\documentstyle[colordvi,prl,aps,psfig,twocolumn]{revtex}
\begin{document}

\def\be{\begin{equation}}
\def\ee{\end{equation}}
\def\ba{\begin{eqnarray}}
\def\ea{\end{eqnarray}}

\title{Density Matrix of a Bose--Einstein Condensate: Steady--State
versus Mean--Field Approach}
\author{S. J. Wang$^1$, M. C. Nemes$^2$, A. N. Salgueiro$^3$,
and H. A. Weidenm\"uller$^4$}
\address{$^1$ Department of Modern Physics, University of Lanzhou,
  730000 Lanzhou, P.R. China}
\address{$^2$Department of Physics, University of Belo Horizonte, Belo
  Horizonte, Minas Gerais, Brasil}
\address{$^3$Max-Planck-Institut f\"ur Physik komplexer Systeme
, D-01187 Dresden, Germany}
\address{$^4$Max-Planck-Institut f\"ur Kernphysik, D-69029 Heidelberg,
  Germany}

\date{\today}

\maketitle

\begin{abstract}
We compare the equilibrium solution for the condensate obtained in the
mean--field approximation to the master equation for sympathetic
cooling with the one obtained by Scully for a system in contact with a
heat bath with the help of an analogy with the laser. While the
mean--field approach yields analytical formulas for the approach towards
equilibrium and for the equilibrium solution, it neglects the
correlations between occupation numbers of different single--particle
states which are approximately kept in Scully's approach. Such neglect
is admissible as long as the fraction of Bosons in the condensate does
not exceed a few percent or so.
\end{abstract}

\pacs{PACS numbers: 05.45.+b, 03.65.Sq, 41.20.Bt, 41.20.Jb}

{\it Motivation.} The experimental realization of Bose--Einstein
condensation~\cite{BCE1} calls for a thorough theoretical understanding
of the properties of the condensate. An important step in this direction
was recently taken by Scully {\it et al.}~\cite{scu99,koc00}. Using the
analogy between the condensate and a laser, these authors derived the
equilibrium form of the reduced density matrix $\rho_0$ of the
condensate for a system containing a fixed number $N_A$ of Bosons. The
result is obtained as the steady--state solution of the time--dependent
non--equilibrium master equation for an ideal Bose gas in a
three--dimensional harmonic trap coupled to a thermal reservoir. The
particle number $N_A$ serves as an important constraint. The solution
found in Refs.~\cite{scu99,koc00} has the form
\be
\rho_0 = \sum_{N \leq N_A} \frac{1}{Z_{N_A}} \bigg [ N_A (\frac{T}
{T_c})^3 \bigg ]^{N_A - N} \Pi^{N} \ .
\label{eq1}
\ee
Here $\Pi^{N} = | N \rangle_0 \ {_0}\langle N |$ is the projector
onto the lowest single--particle state containing $N \leq N_A$ Bosons.
The temperature of the reservoir is denoted by $T$, while $T_c$ is a
suitably defined transition temperature, and $Z_N$ is a normalization
factor.

Another starting point towards understanding the properties of the 
condensate is the master equation for sympathetic cooling. This master
equation was, in very general form, derived in Ref.~\cite{lew95}. The
cooling agent is a cooled gas in thermal equilibrium. The cooling
mechanisms for a system in contact with a reservoir~\cite{scu99} and
for a system subject to sympathetic cooling~\cite{lew95} differ. In
the first case, energy is exchanged via exciting or de--exciting
the reservoir. In the second case, cooling is due to two--body
collisions between the atoms in the cooling gas and those in the
system of $N_A$ Bosons. The form given in Eq.~(\ref{eq1}) for the
density matrix of the condensate is also obtained from the master
equation for sympathetic cooling~\cite{pap01} by postulating that the
elements of the density matrix referring to excited single--particle
states have attained the equilibrium form but that this form depends
on the number $N$ of Bosons in the ground--state configuration
$| N \rangle_0$. The agreement between the forms of $\rho_0$ found
from two different approaches gives confidence that Eq.~(\ref{eq1})
represents a good approximation to the true from of the density matrix
for the condensate.

Recently, the present authors have developed a mean--field approach
to the master equation for sympathetic cooling~\cite{wan01}. This
approach yields explicit analytic expressions for the density matrix 
of both, the condensate and the $N_A$--Boson system in excited
single--particle states. The approach yields not only the
equilibrium solution but also the time--dependence of the cooling
process. The reason is that the mean--field equations possess a
$SU(1,1)$ dynamical symmetry. Therefore, the equations are
integrable~\cite{wan93,wan89}.

In the present Letter, we compare the equilibrium solution for the
density matrix $\rho_0$ of the condensate as obtained from the
mean--field approximation with Eq.~(\ref{eq1}). The comparison will
cast new light on both, the form of $\rho_0$ and the limitation of the
mean--field approach. In addition, we briefly display the time
evolution towards  equilibrium as obtained in the mean--field approach.

{\it Master Equation.} Starting point is the master equation derived
in Ref.~\cite{lew95}. We use the notation of Refs.~\cite{lew95,pap01}.
We label the system subject to sympathetic cooling as system $A$. This
system consists of $N_A$ Bosons. The master equation for the dependence
of the reduced density matrix $\rho_A(t)$ for system $A$ on time $t$
reads

\begin{equation}
\label{master}
\frac{{\rm d} \rho_A(t)}{{\rm d}t} = - \frac{i}{\hbar} \biggl [ H_A +
H'_{A-A}, \rho_A(t) \biggr ] + {\cal L} \rho_A \ .
\end{equation}

\noindent
Here, $H_A$ is the sum of the single--particle Hamiltonians for the
atoms in system $A$ (each containing the kinetic energy and the
harmonic trap potential) while $H'_{A-A}$ represents the (weak)
interaction between the atoms in system $A$. This interaction is
neglected in what follows because sympathetic cooling is used
precisely when $H'_{A-A}$ is very small. The action of the
Liouvillean $\cal L$ on the reduced density matrix $\rho_A(t)$ is
given by

\begin{eqnarray}
\label{eq2}
{\cal L} \rho_A &=& \sum_{{\vec n}, {\vec n}', {\vec m}, {\vec m}'}
\Gamma^{{\vec m},{\vec m}'}_{{\vec n},{\vec n}'} \biggl ( 2
a^{\dagger}_{\vec m} a_{{\vec m}'} \rho_A(t) a^{\dagger}_{\vec n}
a_{{\vec n}'} 
\nonumber \\
&&- a^{\dagger}_{\vec n} a_{{\vec n}'} a^{\dagger}_{\vec m} a_{{\vec
 m}'} \rho_A(t) - \rho_A(t) a^{\dagger}_{\vec n} a_{{\vec n}'}
a^{\dagger}_{\vec m} a_{{\vec m}'} \biggr ) \ .
\end{eqnarray}

\noindent
The single--particle states of the three--dimensional isotropic
harmonic trap are labelled by the quantum numbers ${\vec m} = (m_x,
m_y, m_z)$ with $m_x, m_y, m_z$ integer. The creation and annihilation
operators for these states are written as $a^{\dagger}_{\vec m}$ and
$a_{\vec m}$, respectively. The index $0$ is used for the
non--degenerate ground state of the trap. The rate coefficients
$\Gamma^{{\vec m}, {\vec m}'}_{{\vec n},{\vec n}'}$ are given in
Ref.~\cite{lew95}. We do not repeat the definition here. Suffice it to
say that these coefficients account for the interaction between
particles in system $A$ and those in the cooling system $B$. We assume
that the number $N_B$ of atoms in system $B$ is very large. Then
decoherence acts very quickly~\cite{pap01} and reduces the density
matrix to diagonal form. Hence,

\ba
{\cal L} \rho_A &=& \sum_{{\vec m} \neq {\vec n}}
\Gamma^{{\vec m},{\vec n}}_{{\vec n},{\vec m}} \biggl ( 2
a^{\dagger}_{\vec m} a_{{\vec n}} \rho_A(t) a^{\dagger}_{\vec n}
a_{{\vec m}} 
\nonumber \\
&&- a^{\dagger}_{\vec n} a_{{\vec m}} a^{\dagger}_{\vec
  m} a_{{\vec n}} \rho_A(t) - \rho_A(t) a^{\dagger}_{\vec n}
a_{{\vec m}} a^{\dagger}_{\vec m} a_{{\vec n}} \biggr ) \ .
\label{eq3}
\ea

{\it Mean--Field Approximation.} Eq.~(\ref{eq3}) serves as the
starting point for the mean--field approximation. We apply this
approximation in standard fashion by replacing on the right--hand side
of Eq.~(\ref{eq3}) one pair of creation and annihilation operators
referring to the same single--particle state ${\vec m}$ or ${\vec n}$
by its expectation value $\langle a_{\vec m}^{\dagger} a_{\vec m}
\rangle$. The quantity

\be
\label{eq3a}
N_{\vec m} = \langle a_{\vec m}^{\dagger} a_{\vec m} \rangle =
{\rm tr} ( a_{\vec m}^{\dagger} a_{\vec m} \rho_A )
\ee

\noindent
is the mean particle occupation number of the single--particle state
${\vec m}$. For the term ${\cal L} \rho_A $, this procedure yields

\ba
\label{eq4}
{\cal L} \rho_A &=& \sum_{{\vec m} \neq {\vec n}}
\Gamma^{{\vec m},{\vec n}}_{{\vec n},{\vec m}} \ N_{\vec n} \ \biggl
( 2 a^{\dagger}_{\vec m} \rho_A(t) a_{\vec m} \nonumber \\
&&-  a^{\dagger}_{\vec m} a_{\vec m} \rho_A(t) - \rho_A(t) a_{\vec m}
a^{\dagger}_{\vec m} - \rho_A \biggr ) \nonumber \\
&&+ \sum_{{\vec m}}\Gamma^{{\vec n}{\vec m}}_{{\vec m}{\vec n}} \
(N_{\vec n} + 1) \ \biggl ( 2 a_{\vec m} \rho_A(t) a^{\dagger}_{\vec m}
\nonumber \\
&&-  a^{\dagger}_{\vec m} a_{\vec m} \rho_A(t) - \rho_A(t) a_{\vec m}
a^{\dagger}_{\vec m} + \rho_A \biggr ) \ .
\ea

\noindent
Central to the mean--field approach is the assumption that there are
no correlations between the occupation probabilities of the harmonic
trap levels $\vec{n}$ and $\vec{m}$. 

{\it Rate Equation.} Eqs.~(\ref{master}) and (\ref{eq4}) can be reduced
to a rate equation. We use a basis in Hilbert space defined by a
product of all single--particle states ${\vec m}$, each such state
being occupied by $N_{\vec m}$ Bosons. We take the trace of
$\rho_{A}(t)$ over all single--particle states $\vec{n}$ with $\vec{n}
\neq \vec{m}$ (this includes a summation over all occupation numbers
$N_{\vec m}$) and denote the result by $\rho_{\vec m}(t)$. That same
notation was used already in Eq.~(\ref{eq1}) with ${\vec m} = 0$. We
recall that $\rho_{A}(t)$ is diagonal in energy representation. It
follows that $\rho_{\vec m}(t)$ can be written as a sum of the
projectors $\Pi^N_{\vec m}=| N \rangle_{\vec m} {_{\vec m}} \langle
N |$,

\be
\label{eq18}
\rho_{\vec m}(t) = \sum_{N = 0}^{N_A} P^N_{\vec m}(t) \ \Pi^N_{\vec m}
\ .
\ee

\noindent
The time--dependent mean--field occupation probabilities
$P^N_{\vec m}(t)$ differ from zero only for $N \leq N_A$. Taking the
corresponding trace of Eq.~(\ref{master}) and using Eq.~(\ref{eq4}),
we find that the $P^N_{\vec m}(t)$ obey the rate equation

\ba
\label{eq19}
\frac{{\rm d} P^N_{\vec m}(t)}{{\rm d}t} &=& 2 K_{\vec{m}} N
P^{N-1}_{\vec m} + 2 H_{\vec m} (N+1) P^{N+1}_{\vec m} \nonumber \\
&&- 2(K_{\vec m} + H_{\vec m}) N P^N_{\vec m} - 2 K_{\vec m}
P^N_{\vec m} \ .
\ea

\noindent
The cooling and heating coefficients $K_{m}$ and $H_{m}$ are given by

\ba
\label{eq20}
&&K_{\vec m} = \sum_{{\vec n} \neq {\vec m}} \Gamma^{{\vec m},{\vec
    n}}_{{\vec n},{\vec m}} N_{\vec n} \ , \nonumber \\
&&H_{\vec m} = \sum_{{\vec n} \neq {\vec m}} \Gamma^{{\vec n},{\vec
    m}}_{{\vec m},{\vec n}} (N_{\vec n} + 1) \ .
\ea

\noindent
>From Eqs.~(\ref{eq3a},\ref{eq18}) and (\ref{eq19}) we obtain for the
mean occupation number $N_{\vec{m}}(t)$

\be
\label{eq21}
\frac{{\rm d}N_{\vec{m}}}{{\rm d}t} = 2 K_{\vec{m}} \ (N_{\vec{m}}+1)
- 2 H_{\vec{m}} \ N_{\vec{m}} \ .
\ee

\noindent
This equation is consistent with conservation of particle number,
$\sum_{\vec m} {\rm d}N_{\vec{m}}/{\rm d}t = 0$.

{\it Condensate.} Putting $\vec{m}=0$ in Eq.~(\ref{eq19}), one
obtains the equation for the probability distribution of the number
of atoms in the single--particle ground state, i.e., the condensate.
The equilibrium solution obeys ${\rm d}P_{0}^{N}(t)/{\rm d}t = 0$.
It is easily seen that the resulting equations for the
time--independent coefficients $P_{0}^{N}$ do not possess a
non--trivial solution. This is because the constraint $P_{0}^{N} =
0$ for $N > N_A$ is too rigid for the mean--field approach. We
relax this condition, write

\be
\rho_{0}(t) = \sum_{N = 0}^{\infty} P^N_{0}(t) \
\Pi^N_{\vec m}
\label{eq01}
\ee

\noindent
and require only that the sum $\sum_{N = N_A}^{\infty} P^N_{0}$
be negligibly small. The resulting equation for the $P^N_{0}$'s reads

\ba
&&2 K_0 N P^{N-1}_0 + 2 H_0 (N+1) P^{N+1}_0 - 2 (H_0 + K_0) N P^N_0
\nonumber \\
&&\qquad \qquad  - 2 K_0 P^N_0 = 0 \ , \qquad N = 0,\ldots,\infty
\ .
\label{eq02}
\ea

\noindent
The solution is

\be
\label{eq24}
P^N_{0} = P^0_{0} \ \chi^N \ \ {\rm where} \ \ \chi =
\frac{K_{0}}{H_{0}} \ .
\ee

\noindent
Eq.~(\ref{eq21}) implies that in the stationary case we must have
$\chi < 1$. The normalization condition yields $P^0_{0} = (1 - \chi)$.
The mean value $N_{0}$ is given by $N_{0} = \chi / (1 - \chi)$.
Conversely, we may replace $\chi$ everywhere by $N_{0}/(N_{0}+1)$. To
discuss the validity of the mean--field solution, we impose the
constraint that $\sum_{N = N_A}^{\infty} P^{N}_{0}/P^{0}_{0} =
\chi^{N_A}/(1 - \chi) < \exp(-a)$. For $N_A \gg 1$, this yields
$N_{0}/N_A < 1/a$. This condition applies as long as $a \gg \ln (N_0)$
and shows that for the condensate, the mean--field approximation (as
defined in the framework of this paper) begins to fail whenever the
ratio $N_{0}/N_A$ grows beyond a few percent or so. Similar conclusions
apply to the stationary occupation probabilities $P^N_{\vec m}$ of
excited single--particle states, except that here we do not expect the
mean occupation number ever to approach values close to $N_A$.

The $P^N_{0} \sim \chi^N$ given by Eq.~(\ref{eq24}) decrease
monotonically with increasing $N$. This behavior is in marked contrast
to that of the equilibrium solution Eq.~(\ref{eq1}) found first by
Scully~\cite{scu99,koc00} and reproduced, in the present context, in
Ref.~\cite{pap01}. The ground--state distribution as given by
Eq.~(\ref{eq1}) is Poissonian, except for the additional normalization
factor.

The physical situations encapsulated in Eqs.~(\ref{eq1}) and
(\ref{eq24}) are completely analogous to the behavior of a laser far
above and below threshold, respectively~\cite{scubook}. The difference
between Eqs.~(\ref{eq1}) and (\ref{eq24}) is due to the strong
correlations between the numbers of particles in excited
single--particle states and in the ground state. That correlation is
taken into account approximately in the derivation of Eq.~(\ref{eq1}).
Indeed, in this derivation it is assumed that for ${\vec m} \neq 0$ we
have $\langle a_{\vec m}^{\dagger} a_{\vec m} \rangle =
N_{\vec{m}}(N)$ with $N_{\vec{m}}(N)$ constrained by $N$, the number
of Bosons in the ground--state configuration $| N \rangle_0$. The
resulting nonlinearity is analogous to the nonlinearity which governs
the behavior of a laser far above threshold. It carries the solution
beyond the mean--field approximation and is well suited to describe
the probability distribution of the atoms in the condensate whenever
their number is of the order of $N_{A}$. As mentioned in the beginning,
there is good reason to believe that Eq.~(\ref{eq1}) correctly
describes this situation. In the mean--field approach, on the other
hand, the correlations between the numbers of particles in the excited
single--particle states and in the ground state are totally neglected.
This situation is analogous to the behavior of a laser below threshold
(linear case). We conclude that Eqs.~(\ref{eq1}) and (\ref{eq24})
describe two different regimes. The equilibrium solution
Eq.~(\ref{eq24}) describes a system where at most a few percent of the
atoms are in the ground state. As that fraction increases, there is a
gradual transition to another regime described by Eq.~(\ref{eq1}). The
intermittent case is apparently not covered by either formula.

In the remainder of the paper, we demonstrate that the mean--field
approximation is well suited to describe the approach of the system
towards equilibrium, and not only the stationary case. This is due to
the fact that the master equation is linear in the generators of the
group $SU(1,1)$. Details of the calculation are given
elsewhere~\cite{wan01}. Writing the time--dependent master equation
in the form

\be
\frac{{\rm d}\rho_{\vec{m}}(t)}{{\rm d}t} = \Gamma(t)
\rho_{\vec{m}}(t) \ ,
\label{eq03}
\ee

\noindent
with $\Gamma(t)$ dependent upon time, we construct a time--dependent
similarity transformation $T(t)$ which diagonalizes $\Gamma(t)$.
Because of the form of $\Gamma$, there exist two different similarity
transformations which accomplish this aim. However, only one of the
two fulfills the condition for the viability of the mean--field
solution, namely, that the coefficients multiplying $\Pi^N$ vanish
asymptotically for large $N$. Using this transformation, the
time--evolution of the reduced density matrix can be determined and
is given by

\ba
\label{eq43}
&&\rho_{\vec m}(t) = \exp[ \alpha_{+}(t) K^{+} ] \ \exp[
\alpha_{-}(t) K^{-} ]
\ \nonumber \\
&&\ \times \exp \biggl( \int_0^t {\rm d}\tau [ \gamma(\tau) K^0
- K_{m}(\tau) - H_{m}(\tau)) ] \biggr) \ \rho_{\vec m}(0)
\ea

\noindent
where $\rho_{\vec m}(0)$ is fixed by the initial condition, and
$\gamma(t) = 4 H_{m}(t) \alpha_{+}(t) - 2 [K_{m}(t) + H_{m}(t)]$. The
time--dependent functions $\alpha_{\pm}(t)$ are the solutions of the
differential equations

\ba
\label{eq40}
&&\frac{{\rm d} \alpha_{+}(t)}{{\rm d}t}= 2K_{m} + 2H_{m} \alpha_{+}^2 -
2(K_{m}+ H_{m}) \ \alpha_{+} \nonumber \\ 
&&\frac{{\rm d} \alpha_{-}(t)}{{\rm d}t} = 2H_{m}
\ (1 - 2 \alpha_{+} \alpha_{-}) + 2(K_{m} + H_{m}) \ \alpha_{-} \ .
\ea

\noindent
The initial conditions are $\alpha_{\pm}(0) = 0$. The action of the
operators $K^{+}$, $K^{-}$ and $K^{0}$ on the quantities
$\Pi^{n,k}_{\vec{m}} = |n \rangle_{\vec{m} \vec{m}}\langle k|$ is
given by

\ba
\label{eqA2}
&&K^{0}_{\vec m} \Pi^{n,k}_{\vec m} = \frac{1}{2}(n+k+1)
\Pi^{n,k}_{\vec m} \ , \nonumber \\
&&K^{+}_{\vec m} \Pi^{n,k}_{\vec m} = \sqrt{(n+1)(k+1)} \;
\Pi^{n+1,k+1}_{\vec m} \ ,
\nonumber \\
&&K^{-}_{\vec m} \Pi^{n,k}_{\vec m} = \sqrt{nk} \;
\Pi^{n-1,k-1}_{\vec m} \ .
\ea

\noindent
>From the asymptotic behavior of the coefficients $K_{m}$, $H_{m}$,
$\gamma$ and $\alpha_{\pm}$, it can be shown that as $t \rightarrow
\infty$, $\rho_{\vec m}(t)$ approaches the equilibrium solution
which for ${\vec m} = 0$ is displayed in Eq.~(\ref{eq24}).

In summary, our results show that the mean--field approach is a useful
tool to investigate both sympathetic cooling and properties of the
condensate. The limiting factor in the approach is the neglect of
correlations between occupation numbers in the ground state and in
excited single--particle states. Such correlations generate a
non-linear term in the cooling and heating coefficients and become
important whenever the fraction of Bosons in the condensate exceeds
a few percent or so. While the master equation for the mean--field
approach allows for an analytical solution, it is unlikely that such
a solution exists in the general case. The distributions of
occupation probabilities of states with $N_0$ Bosons in the ground
state differ very much in the mean--field approach and in the
approximation obtained by Scully. The latter is probably adequate
whenever the fraction of Bosons in the ground state is of order unity.
The two distributions display similarities with the behavior of a
laser below and far above threshold, respectively.

{\it Acknowledgment.} This work was completed while one of us (HAW)
participated in the workshops on ``Nanostructures'' and on ``Quantum
Information'' held by the ITP at UCSB. He is grateful to the organizers
for the invitation, and for financial support given by National Science
Foundation grant number PHY 99-07949. ANS acknowledges the
Max Planck Institute of Nuclear Physics at Heidelberg where this work started
and also the support given by FAPESP (Funda\c{c}\~ao de Amparo a Pesquisa
do Estado de S\~ao Paulo) at that time. The Author WSJ is 
grateful to the Max Planck 
Institute of Nuclear Physics at Heidelberg, the Max Planck Gemeinschaft(MPG), 
and the National Natural Science Foundation of China for their 
financial support. MCN acknowledges Conselho Nacional de Desenvolimento
Cient\'{\i}fico e Tecnol\'ogico (CNPq), the Max Planck Institute of Nuclear 
Physics at Heidelberg and the Humboldt Foundation.


\begin{thebibliography}{99}

\bibitem{BCE1} M. Anderson, J. R. Ensher, M. R. Matthews, C. E. Wieman
  and E. A. Cornell, Science {\bf 269}, 198 (1995); C. Bradley, C.
  Sackett, J. Tollett, and R. Hulet, Phys. Rev. Lett. {\bf 75}, 1687
  (1995); K. Davis, M. Mewes, M. R. Andrews, N. van Druten, D. S. Durfee,
  D. Kurn, and W. Ketterle, Phys. Rev. Lett. {\bf 75}, 3969 (1995).
\bibitem{scu99}M. O. Scully, Phys. Rev. Lett. {\bf 82} (1999) 3927.
\bibitem{koc00}V. V. Kocharovsky, M. O. Scully, S.-Y. Zhu, and
  M. S. Zubairy, Phys. Rev. {\bf A 61} (2000) 023609.
\bibitem{lew95}M. Lewenstein, J. I. Cirac, and P. Zoller,
  Phys. Rev. {\bf A 51} (1995) 4617.
\bibitem{pap01}T. Papenbrock, A. N. Salgueiro, and
  H. A. Weidenm\"uller, cond-mat/0106392 and cond-mat/01006418.
\bibitem{wan01}S. J. Wang, M. C. Nemes, A. N. Salgueiro and
  H. A. Weidenm\"uller, cond-mat/0110293.
\bibitem{wan93}S. J. Wang, F. L. Li, and A. Weiguny, Phys. Lett. {\bf
  A 180} (1993) 189.
\bibitem{wan89}S. J. Wang, J. M. Cao, and A. Weiguny, Phys. Rev. {\bf
  A 40} (1989) 1225.
\bibitem{scubook} M. O. Scully and M. Suhail Zubairy, {\it Quantum
  Optics} Cambridge Uni. Press, Cambridge (1999).





\end{thebibliography}
\end{document}